\documentclass[graybox]{svmult}

\usepackage{mathptmx}      
\usepackage{helvet}       
\usepackage{courier}       
\usepackage{type1cm}        
\usepackage{makeidx}         
\usepackage{graphicx}        
\usepackage{multicol}        
\usepackage[bottom]{footmisc}

\usepackage{mathtools}

\makeindex

\begin{document}

\title*{Thirty eight things to do with live slime mould\thanks{This is a draft version of the chapter to appear in Advances in Unconventional Computing, Springer, 2016.}}
\titlerunning{Thirty eight things to do with live slime mould} 
\author{Andrew Adamatzky}
 \authorrunning{Andrew Adamatzky}
 \institute{Andrew Adamatzky \at  Unconventional Computing Centre, University of the West of England, Bristol, UK\email{andrew.adamatzky@uwe.ac.uk}}

\maketitle

\markboth{Adamatzky A. Thirty eight things to do with live slime mould. December 2015}{Adamatzky A. Thirty eight things to do with live slime mould. December 2015}

\abstract{Slime mould \emph{Physarum polycephalum} is a  large single cell capable for distributed sensing, concurrent information processing, parallel computation and decentralised actuation. The ease of culturing and experimenting with Physarum makes this slime mould  an ideal substrate for real-world implementations of unconventional sensing and computing devices. In the last decade the Physarum became a swiss knife of the unconventional computing: give the slime mould a problem it will solve it. We provide a concise summary of what exact computing and sensing operations are implemented with live slime mould. The Physarum devices range from morphological processors for computational geometry to experimental archeology tools, from self-routing wires to memristors, from devices approximating a shortest path to analog physical models of space exploration. }

\abstract*{Slime mould \emph{Physarum polycephalum} is a  large single cell capable for distributed sensing, concurrent information processing, parallel computation and decentralised actuation. The ease of culturing and experimenting with Physarum makes this slime mould  an ideal substrate for real-world implementations of unconventional sensing and computing devices. In the last decade the Physarum became a swiss knife of the unconventional computing: give the slime mould a problem it will solve it. We provide a concise summary of what exact computing and sensing operations are implemented with live slime mould. The Physarum devices range from morphological processors for computational geometry to experimental archeology tools, from self-routing wires to memristors, from devices approximating a shortest path to analog physical models of space exploration.}

\tableofcontents

\newpage

\markboth{Adamatzky A. Thirty eight things to do with live slime mould. December 2015}{Adamatzky A. Thirty eight things to do with live slime mould. December 2015}

\section*{Introduction}

Acellular slime mould \emph{P. polycephalum} has quite sophisticated life cycle~\cite{stephenson1994myxomycetes}, which includes fruit bodied, spores, single-cell amoebas, syncytium. At one phase of its cycle the slime mould becomes a plasmodium. The plasmodium is a coenocyte: nuclear divisions occur without cytokinesis. It is  a single cell with thousands of nuclei. The plasmodium is a large cell. It grows up to tens centimetres when conditions are good.
The plasmodium consumes microscopic particles and bacteria. During its  foraging behaviour the plasmodium spans scattered sources of nutrients with a network of  protoplasmic tubes. The plasmodium optimises it protoplasmic network to cover all sources of nutrients, stay away from repellents and minimise transportation of metabolites inside its body. The plasmodium's ability to optimise its shape~\cite{nakagaki2001path} attracted attention of biologists, then computer scientists~\cite{adamatzky2010physarum}  and engineers. Thus the field of slime mould computing was born. 

So far, the plasmodium is the only useful for computation stage of \emph{P. polycephalum}'s life cycle. Therefore further we will use word `Physarum' when referring to  the plasmodium. Most computing and sensing devices made of the Physarum explore one or more key features of the Physarum's physiology and behaviour: 
\begin{itemize}
\item the slime mould senses gradients of chemo attractants and repellents~\cite{durham1976control, ueda1976chemotaxis, rakoczy2015application}; it responds to chemical or physical stimulation by changing patterns of  electrical potential oscillations~\cite{ridgway1976oscillations, kishimoto1958rhythmicity} and protoplasmic tubes contractions~\cite{wohlfarth1979oscillatory, teplov1991continuum};  
\item it optimises its body to maximise its protoplasm streaming~\cite{dietrich2015explaining}; and, 
\item it is made of hundreds, if not thousands, of biochemical oscillators~\cite{kauffman1975mitotic} with varied modes of coupling~\cite{grebecki1978plasmodium}.  
\end{itemize}
Here we offer very short descriptions of actual working prototypes of Physarum based sensors, computers, actuators and controllers. Details can be found in pioneer book on Physarum machines~\cite{adamatzky2010physarum}  and the `bible' of slime mould computing~\cite{adamatzkyAdvancesPhysarum}.


\section{Optimisation and graphs}

\subsection{Shortest path and maze\index{maze}}
\label{path}

Given a maze we want to find a shortest path between the central chamber and an exit. 
This was the first ever problem solved by Physarum. There are two Physarum processors which solve the maze. First prototype~\cite{nakagaki2001path} works as follows. The slime mould is inoculated everywhere in a maze. The Physarym develops a network of protoplasmic tubes spanning all channels of the maze. This network represents all possible solutions. Then oat flakes are placed in a source and a destination site. Tube lying along the shortest (or near shortest) path between two sources of nutrients develop increased flow of cytoplasm. This tube becomes thicker. Tubes branching to sites without nutrients become smaller due to lack of cytoplasm flow. They eventually collapse. The sickest tube represents the shortest path between the sources of nutrients. The selection of the shortest protoplasmic tube is implemented via interaction of propagating bio-chemical, electric potential and contractile waves in the plasmodium's body, see mathematical model in~\cite{tero2006physarum}.  The approach is not efficient because we must literally distribute the computing substrates everywhere in the physical representation of the problem. A number of computing elements would be proportional to a sum of lengths of the maze's channels.
\index{shortest path}

Second prototype of the Physarum maze solver is based on Physarum' chemo-attraction~\cite{adamatzky2012slimemaze}. An oat flake is placed in the central chamber. The Physarum is inoculated somewhere in in a peripheral channel. The oat flake releases chemoattractants. The chemoattractants diffuse along the maze's channels. The Physarum explores its vicinity by branching out protoplasmic tubes into  opening of nearby channels.  When a wave-front of diffusing attractants reaches Physarum, the Physarum halts lateral exploration. Instead it develops an active growing zone propagating along gradient of the  attractants' diffusion. The sickest tube represents the shortest path between the sources of nutrients. The approach is efficient because a number of computing elements would be proportional to a length of the shortest path.

\subsection{Towers of Hanoi\index{Towers of Hanoi}}

Given $n$ discs, each of unique size,  and three pegs, we want to move the entire stack to another peg by moving one top disk  at a time and not placing a disk on top of smaller disc. The set of all possible configurations and moves of the puzzle forms a planar graph with $3n$ vertices.  To solve the puzzle one must find shortest paths between configurations of pegs with discs on the graph~\cite{hinz1989tower, hinz1992shortest, romik2006shortest}.  Physarum solves shortest path \index{shortest path} (Sect.~\ref{path}) therefore it can solve Tower of Hanoi puzzle. This is experimentally demonstrated in \cite{reid2013solving}. Sometimes Physarum does not construct an optimal path initially.  However, if its protoplasmic networks are damaged and then allowed to regrow the path closer to optimal then before develops~\cite{reid2013solving}. 

\subsection{Travelling salesman problem\index{Travelling salesman problem}}

Given a graph with weighted edges, find a cyclic route on the graph, with a minimum sum of edge weights, spanning all nodes, where each node is visited just once. Commonly, weight of an edge is an Euclidean length of the edge. 
Physarum is used as component of an experimental device approximating the shortest cyclic route~\cite{zhu2013amoeba}. A map of eight cities is considered. A set of solutions is represented by channels arranged in a star graph.  The channels merge in a central chamber. There are eight channels for each city. Each channel encodes a city and the step when the city appears in the route. There are sixty four channels. Physarum is inoculated in the central chamber. The slime mould then propagates into the channels.  A node of the data graph is assumed to be visited when its corresponding channel is colonised by Physarum. Growth of the Physarum in the star-shape is controlled by a recurrent neural network. The network takes ratios of colonizations of the channels as input and produces patterns of illumination, projected onto the channels, as output.  The network is designed to prohibit revisiting of already visited nodes and simultaneous visits to multiple nodes. The  Physarum propagates into channels and pulls back. Then propagates to other channels and pulls back from some of them. Eventually the system reaches a stable solution where no propagation occurs.  The stable solution represents the minimal distance cyclic route on the data graph~\cite{zhu2013amoeba}.  

A rather more natural approximate algorithm of solving the travelling salesman problem is based on aconstruction of an $\alpha$-shape (Sect.~\ref{concavehull}). This how humans solve the problem visually~\cite{macgregor1996human}.
An approximate solution of the travelling salesman problem by a shrinking blob of  simulated Physarum is proposed in \cite{jones2014computation}.  The Physarum is inoculated all over the convex hull of the data set. The Physarum's blob shrinks. It adapts morphologically to the configuration of data nodes. The shrinkage halts when the Physarum no longer covers all data nodes. The algorithm is not implemented with real Physarum.

\subsection{Spanning tree\index{spanning tree}\index{tree!spannin}}

A spanning tree of a finite planar set is a connected, undirected, acyclic planar graph, whose vertices
are points of the planar set. The tree is a minimal spanning tree where sum of edge lengths is minimal~\cite{nevsetvril2001otakar}.  As algorithm for computing a spanning tree of a finite planar set based on
morphogenesis of a neuron's axonal tree was initially proposed~\cite{adamatzky1991neural}: planar data points are marked by attractants (e.g. neurotrophins) and a neuroblast is placed at some site. Growth cones sprout new filopodia in the direction of maximal concentration of attractants. If two growth cones compete for the same site of attractants then a cone with highest energy (closest to previous site or branching point) wins. Fifteen years later we implemented the algorithm with Physarum~\cite{adamatzky2008growing}.   

Degree of Physarum branching is inversely proportional to a quality of its substrate. Therefore to reduce a number of random branches we cultivate  Physarum not on agar but just humid filter paper. Planar data set is represented by
a configuration of oat flakes. Physarum is inoculated at one of the data sites. Physarum propagates to a virgin oat flake
closest to the inoculation site. Physarum branches, if there are several virgin flakes nearby. It colonises next set of flakes. The propagation goes on until all data sites are spanned by a protoplasmic network. The protoplasmic network approximates the spanning tree. The resulted tree does not remain static though. Later cycles can be formed and the tree is transformed to one of proximity graphs, e.g. relative neighbourhood graph or Gabriel graph~\cite{adamatzky2009developing}.

\subsection{Approximation of transport networks\index{transport networks}}
\label{roadnetworks}

Motorway networks are designed with an aim of efficient vehicular transportation of goods and passengers. Physarum  protoplasmic networks evolved for efficient intra-cellular transportation of nutrients and metabolites. To uncover similarities between biological and human-made transport networks and to project behavioural traits of biological networks onto development of vehicular transport networks we conducted  an evaluation and approximation of motorway networks by Physarum in fourteen geographical regions:   Africa, Australia, Belgium, Brazil, Canada, China, Germany, Iberia, Italy, Malaysia, Mexico, The Netherlands, UK, and USA~\cite{adamatzky2012bioevaluation}.

We represented each region with an agar plate, imitated major urban areas with oat flakes, inoculated Physarum  in a capital, and analysed structures of protoplasmic networks developed. We found that,  the networks of protoplasmic tubes grown by Physarum match, at least partly, the networks of human-made transport arteries. The shape of a country and the exact spatial distribution of urban areas, represented by sources of nutrients, play a key role in determining the exact structure of the plasmodium network.  In terms of absolute matching between Physarum networks and motorway networks the regions studied can be arranged in the following order of decreasing matching: Malaysia, Italy, Canada, Belgium, China, Africa, the Netherlands, Germany, UK, Australia, Iberia, Mexico, Brazil, USA.  We compared the Physarum and the motorway graphs using such measures as average and longest shortest paths, average degrees, number of independent cycles, the Harary index, the $\Pi$-index and the Randi\'{c} index. Using these measures we find that motorway networks in Belgium, Canada and China are most affine to Physarum networks. With regards to measures and topological indices we demonstrated that  the Randi\'{c} index could be considered as most bio-compatible measure of transport networks, because it matches very well the slime mould and man-made transport networks, yet efficiently discriminates between transport networks of different regions~\cite{adamatzky2013motorways}. 

Many curious discoveries have been made. Just few of them are listed below. All segments of trans-African highways not represented by Physarum have components of non-paved roads~\cite{adamatzky2013biological}.  The east coast transport chain from the Melbourne urban area in the south to the Mackay area in the north, and the highways linking Alice Springs and Mount Isa and Cloncurry, are represented by the slime mould's protoplasmic tubes in almost all experiments on approximation of Australian highways~\cite{adamatzky2012slimeAustralia}. If the two parts of Belgium were separated with Brussels in Flanders, the Walloon region of the Belgian transport network would be represented by a single chain from Tournai in the north-west to the Liege area in the north-east and down to southernmost Arlon; motorway links connecting Brussels with Antwerp, Tournai, Mons, Charleroi and Namur, and links connecting Leuven with Liege and Antwerp with Genk and Turnhout, are redundant  from the Physarum's point of view~\cite{adamatzky2012slimeBelgium}. The protoplasmic network forms a subnetwork of the man-made motorway network in the Netherlands; a flooding  of large area around Amsterdam will lead to  substantial increase in traffic at the boundary between flooded and non-flooded area, paralysis and abandonment of the transport network and migration of population from tshe Netherlands to Germany, France and Belgium~\cite{adamatzky2013bioNetherlands}. Physarum imitates the 1947 year separation of Germany into East Germany and West Germany~\cite{adamatzky2012schlauschleimer}.

\subsection{Mass migration\index{mass migration}}
\label{migration}

People migrate towards sources of safe life and higher income.  Physarum migrates into environmentally conformable areas and towards source of nutrients. In~\cite{adamatzky2013bio} we explored this analogy to imitate Mexican migration to USA. We have made a 3D Nylon terrain of USA and placed oat flakes, to act as sources of attractants and nutrients, to ten areas with highest concentration of migrants: New York, Jacksonville, Chicago, Dallas,  Houston, Denver, Albuquerque, Phoenix, Los Angeles, San Jose. We inoculated Physarum in a locus between Ciudad Ju\'{a}rez and Nuevo Laredo, allowed it to colonise the template for five to ten days, and analysed routes of migration.  From results of laboratory experiments we extracted topologies of migratory routes, and highlighted a role of elevations in shaping the human movement networks.

\subsection{Experimental archeology\index{archeology}}

Experimental archeology uses  analytical methods, imaginative experiments, or transformation of a matter~\cite{ascher1961experimental, ingersoll1977experimental, coles1979experimental} in a context of human activity in the past. Knowing that Physarum is fruitful substrate for simulating transport routes (Sect.~\ref{roadnetworks}) and migration (Sect.~\ref{migration}) we explored foraging behaviour of the Physarum to imitate development of Roman roads in the Balkans~\cite{evangelidis2015slime}.  Agar plates were cut in a shape of Balkans area. We placed oat flakes in seventeen areas,  corresponding to Roman provinces and major settlements and inoculated Physarum in Thessaloniki. We found that Physarum imitates growth of Roman roads to a larger extent. For example, the propagation of Physarum from Thessaloniki towards the area of Scopje and Stoboi matches the road aligned with the Valley of Axios, which is  a key communication artery between the Balkan hinterland and Aegean area from Bronze age.  A range of historical scenarios was uncovered~\cite{evangelidis2015slime}, including movement along via Diagonalis, the long diagonal axis that crossed central Balkan; propagation to the East towards Byzantium, towards the North along the coast of Euxeinus Pontus, and from Thessaloniki to Dyrrachion, along the western part of via Egnatia~\cite{evangelidis2015slime}.

 \subsection{Evacuation\index{evacuation}}

Evacuation is a rapid but temporary removal of people from the area of danger. Physarum moves away from sources of repellents or areas of uncomfortable environmental conditions. Would Physarum be able to find a shortest route of evacuation in geometrically constrained environment? To find out we undertook a series of experiments~\cite{kalogeiton2015cellular}. We made a physical, scaled down, model of a whole floor of the real office building and inoculated Physarum in one of the rooms. In first scenario we placed a crystal of a salt in the room with Physarum in a hope that a gradient of sodium chloride diffusion would repel Physarum towards exist of the building along the shortest path. This did not happen. Physarum got lost in the template.  By placing attractants near the exit we rectified the mishap and allowed Physarum to find a shortest route away from the `disaster'~\cite{kalogeiton2015cellular}.  Evacuation is one of few problems where computer models of Physarum find better solution than the living Physarum~\cite{kalogeiton2015biomimicry}.

 \subsection{Space exploration\index{space exploration}}

We employed the foraging behaviour of the Physarum to explore scenarios of future colonisation of the Moon and the Mars~\cite{adamatzky2014slime}. We grown Physarum on three-dimensional templates of these planet and analysed formation of the exploration routes, dynamical reconfiguration of the transportation networks as a response to addition of hubs.  The developed infrastructures were explored using proximity graphs and Physarum inspired algorithms of supply chain designs. Interesting insights about how various lunar missions will develop and how interactions between hubs and landing sites can be established are given in ~\cite{adamatzky2014slime}.

\section{Geometry}


\subsection{Voronoi diagram}
\label{voronoi}

Let $\mathbf P$ be a non-empty finite set of planar points. A planar Voronoi diagram \index{Voronoi diagram} of the set $\mathbf P$ is a partition of the plane into such regions that, for any element of $\mathbf P$, a region corresponding to a unique point $p$ contains all those points of the plane which are closer to $p$ than to any other node of $\mathbf P$. A unique region $vor(p) = \{z \in {\mathbf R}^2: d(p,z) < d(p,m)\, \forall m \in {\mathbf R}^2, \, m \ne z \}$ assigned to the point $p$ is called a Voronoi cell of the point $p$. The boundary of the Voronoi cell of the
  point $p$ is built of segments of bisectors separating pairs of geographically closest points  of the given planar set $\mathbf P$. A union of all boundaries of the Voronoi cells determines the  planar Voronoi diagram: $VD({\mathbf P}) = \cup _{p \in {\mathbf P}} \partial vor(p)$~\cite{preparata1985computational}.
  
 The basic concept of constructing Voronoi diagrams with reaction--diffusion systems is based on an intuitive technique for detecting the bisector points separating two given points of the set $\mathbf P$. If we drop reagents at the two data points the diffusive waves,  or phase waves if the computing substrate is active, travel outwards from the drops. The waves travel the same distance
from the sites of origin before they meet one another. The points where the waves meet are the bisector points~\cite{adamatzky2005reaction}.  

Plasmodium growing on a nutrient substrate from a single site of inoculation expands circularly as a typical diffusive or excitation wave. When two plasmodium waves encounter each other, they stop propagating. To approximate a Voronoi diagram with Physarum~\cite{adamatzky2010physarum}, we physically map a configuration of planar data points by inoculating plasmodia on a substrate. Plasmodium waves propagate circularly from each data point  and stop when they collide with each other. Thus, the plasmodium waves approximate a Voronoi diagram, whose edges are the substrate's loci not occupied by plasmodia. Time complexity of the Physarum computation is proportional to a maximal distance between two geographically neighbouring data points, which is capped by a diameter of the data planar set, and does not depend on a number of the data points.

\subsection{Delaunay triangulation}

Delaunay triangulation is a  dual graph of Voronoi diagram. \index{Delaunay triangulation}
 A Delaunay triangulation of a planar set is a triangulation of the set such that  a circumcircle of any triangle does not contain a point of the set~\cite{delaunay1934sphere}.    There are two ways to approximate the Delaunay triangulation with Physarum.  First is  based on the setup of Voronoi diagram processor (Sect.~\ref{voronoi}). Previously, we wrote  that when propagating fronts of Physarum meet they stop. This is true. But what happens after they stop is equally interesting. Physarum forms a bridge --- a single protoplasmic tube ---- connecting the stationary Physarum fronts. Such protoplasmic tubes typically span geographically neighbouring sites of inoculation and they cross sites of first contacts of growing wave fronts.  These tubes represent edges of the Delaunay triangulation. This is why we proposed in \cite{shirakawa2009simultaneous} that the Voronoi diagram and the Delaunay triangulation are constructed simultaneously by Physarum growing on a nutrient agar.   

Second method of approximating the Delaunay triangulation is implemented on a non-nutrient agar. We represent planar data points by inoculation sites. The inoculants propagate and form a planar proximity graph spanning all inoculation sites.  At the beginning of such development a Gabriel graph~\cite{gabriel1969new} is formed. \index{Gabriel graph} Then additional protoplasmic tubes emerge and the Gabriel graph is transformed to the Delaunay triangulation~\cite{adamatzky2009developing}.

\subsection{Concave hull}
\label{concavehull}

The $\alpha$-shape of a planar set $\mathbf P$  is an intersection of the complement of all closed discs of radius $1/\alpha$ \index{hull} \index{$\alpha$-shape}
that includes no points  of $\mathbf P$~\cite{edelsbrunner1983shape}.  A concave hull is a connected \index{concave hull}
$\alpha$-shape without holes. This is  a non-convex polygon representing area occupied by  $\mathbf P$. 
Given  planar set $\mathbf P$ represented by physical objects  Physarum must approximate
concave hull of $\mathbf P$ by its thickest protoplasmic tube. We represent data points by somniferous pills placed 
directly on a non-nutrient agar. The pills emits attractants to `pull' Physarum towards $\mathbf P$ but they also emit
repellents preventing Physarum from growing inside $\mathbf P$~\cite{adamatzky2012slime}. The combination of 
long-distance attracting forces and short-distance (`short' is $O(D)$, where $D$ is a diameter of $\mathbf P$) repelling forces allows us to implement Jarvis wrapping algorithm~\cite{jarvis1973identification}. We select a starting point which is extremal point of $\mathbf P$. We pull  a rope to other extremal point. We continue until the set $\mathbf P$ is wrapped  completely. We tested feasibility of the idea in laboratory experiments~\cite{adamatzky2012slime}.  In each experiment we arranged several half-pills in a random fashion near centre of a Petri dish and inoculated an oat flake colonised by Physarum few centimetres away from the set $\mathbf P$.  Physarum propagates towards set $\mathbf P$ and starts enveloping the set with its body and the network of protoplasmic tubes. The plasmodium does not propagate inside configuration of pills. The plasmodium completes approximation of a shape by entirely enveloping $\mathbf P$ in a day or two.

\section{Computing circuits}

\subsection{Attraction-based logical gates\index{attraction-based logical gates}\index{logical gate!attraction-based}}

When two growing zones of separate Physarum cells meet they repel if there is a free space to deviate to. If there is no opportunity to deviate the cells merge. This feature is employed in the construction of Boolean logical gates --- {\sc not}, {\sc or} and {\sc and} ---- in \cite{tsuda2004robust}. The gates are made of segments of agar gel along which the Physarum propagates. To implement input `1' ({\sc True}) in channel $x$ a piece of Physarum is inoculated in $x$ otherwise the input is considered to be `0'  ({\sc False}). Attractants are placed in the end of the output channels to stimulate growth of the Physarum towards outputs.  The Physarum propagates towards closest source of attractants along a shortest path.

The gate {\sc or} is a 
$
\begin{smallmatrix}
 	\searrow & 			&  \swarrow	\\
		   & 	\downarrow &	 	
\end{smallmatrix}
$
junction.  Physarum placed in one of the inputs propagates towards the output.  If each input contains the Physarum, the propagating cells merge and appear in the output as if they were a single cell. Even if both inputs are `1' the Physarum cells have no space to avoid collision and therefore the merge and propagate into the output channel.

The gate {\sc and} looks like distorted `H':
$
\begin{smallmatrix}
 		& 	&  \, &  \downarrow	 	& 	&	  	\downarrow\\ \hline
\downarrow	& 	&  \,  & 	& 	 	\downarrow&	  	\\ 
\end{smallmatrix}
$
When only one input is `1' the Physarum propagates towards closes attractant and exits along right output channel. 
When both inputs are `1' the Physarum from the  right input channel propagates into the right output channel. The Physarum from the left input channel avoids merging with another Physarum and propagates towards left output channel.  The left output channel realises {\sc and}.

\subsection{Ballistic logical gates\index{ballistic logical gates}\index{logical gate!ballistic}}
\label{ballistic}

In designs of ballistic gates \cite{adamatzky2010slimeballistic} we employ inertia of the  Physarum growing zones. On a non-nutrient substrate the plasmodium propagates as a traveling localisation, as a compact wave-fragment of protoplasm. The plasmodium-localisation  travels in its originally predetermined direction for a substantial period of time even when  no gradient of chemo-attractants is present.  We explore this feature of Physarum  localisations to design a two-input two-output Boolean gates. The gate realising {\sc and} on one output and {\sc or} on another output look like horizontally flipped `K': 
$
\begin{smallmatrix}
 \searrow & \downarrow \\ 
 \swarrow & \downarrow 
\end{smallmatrix} .
$
When left input is `1' the Physarum propagates inertially along the vertical (on the right) output channel. The same happens when right input is `1'.  If both inputs are `1' then the Physarum from the right input propagates along vertical output channel but the Physarum from the left input repels from the right-input-Physarum and moves into the left output channel.   The left output channel realises {\sc and} and the right output channel realises {\sc or}. 

The gate {\sc not} is an asymmetric cross junction: 
$
\begin{smallmatrix}
    		& |    			&  \\ 
     		& \downarrow    &  \\ 
 \rightarrow & 		 &	\rightarrow  \\
 		& \downarrow 
\end{smallmatrix} .
$
Vertical input channel is twice as long as horizontal input channel.
Vertical input is constant {\sc True}: Physarum is always inoculated their. 
Horizontal input is a variable.  
If variable input is `0' then Physarum from the constant {\sc True} vertical input propagates into the vertical output. 
If variable input is `1' then Physarum from the input channel propagates into the horizontal output channel and blocks  path of  the Physarum representing constant  {\sc True}. Both ballistic gates work very well without attractants. However they work ever better when attractants are placed into output channels. Cascading of the gates into a binary adder is demonstrated in \cite{adamatzky2010slimeballistic}.

\subsection{Opto-electronics logical gates\index{opto-electronics logical gates}\index{logical gate!opto-electronics}}

In prototypes of repellent gates~\cite{mayne2015slimegates}, active growing zones of slime mould representing different inputs interact  with each other by electronically switching light inputs and thus invoking photo avoidance. The gates {\sc not} and {\sc band} are constructed using this feature. 

The gate {\sc not} is made of two electrodes. The electrodes are connected to a power supply. There is a green LED (with it is independent power supply) on one electrode. The Physarum is inoculated on another electrode. Input `1' is represented by LED's light on. Output is represented by the Physarum closing the circuit between two electrodes. When LED is illuminated (input `1') the Physarum does not propagate between electrodes, thus output '0' is produced. When LED is off (input `0') the Physarum closes the circuit by propagating between the electrodes. 
 
 The gate {\sc nand} is implemented with  two LEDs.  When both inputs are `0' LEDs are off and the Physarum closes the circuit between its inoculation electrode and one of the LED electrodes, chosen at random.  When both inputs are `1', both LEDs are on, they repel Physarum. The Physarum does not propagate to any electrodes. When only one input is `1', one LED is on and another LED is off, the Physarum propagates toward the electrode with non-illuminating LED and closes the circuit.

\subsection{Frequency based logical gates\index{frequency based logical gates}\index{logical gates!frequency based}}

The Physarum responds to stimulation with light, nutrients and heating by changing frequency of its electrical potential oscillations. We represent {\sc True} and {\sc False} values by different types of stimuli and apply threshold operations to frequencies of the Physarum oscillations. We represent Boolean inputs as intervals of oscillation frequency~\cite{whiting2014slimefrequency}. Thus we experimentally implement {\sc or}, {\sc and}, {\sc not}, {\sc nor}, {\sc nand}, {\sc xor} and {\sc xnor} gates, see details in  \cite{whiting2014slimefrequency}.

\subsection{Micro-fluidic logical gates\index{micro-fluidic logical gates}\index{logical gates!micro-fluidic}}

When a fragment of protoplasmic tube is mechanically stimulated, e..g. gently touched by a hair,  
a cytoplasmic flow in this fragment halts and the fragment's resistivity to the flow dramatically increases. 
The cytoplasmic flow is then directed via adjacent protoplasmic tubes. A basic gate looks like a `Y'-junction of two protoplasmic tubes $x$ and $y$ with a horizontal bypass $z$ between them
$$\includegraphics[scale=0.3]{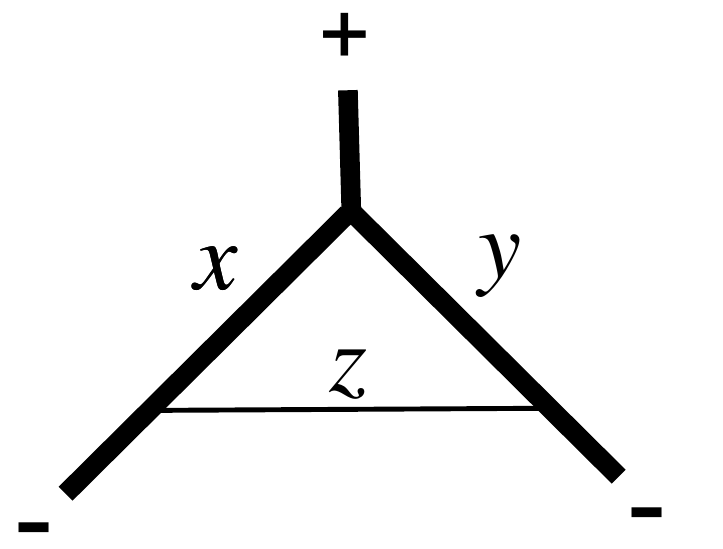}.$$ Segments $x$ and $y$ are inputs. Segment $z$ is output. 
When both input segments are intact and there is a flow of cytoplasm between them there is no flow 
of cytoplasm via $z$.  If one of the input tubes is mechanically stimulated the flow through this tube stops and the flow is diverted via the output tube $z$. If both input tubes are stimulated there is no flow via input or output tubes.  Thus we implement {\sc xor} gate~\cite{adamatzky2014slimefluidic}. A mechanically stimulated 
fragment restores its flow of cytoplasm in one minute: the gate is reusable. 

By adding one more output (bypass) tube to {\sc xor} gate we produce a gate with two inputs and two outputs: $z$ and $p$:  $$\includegraphics[scale=0.3]{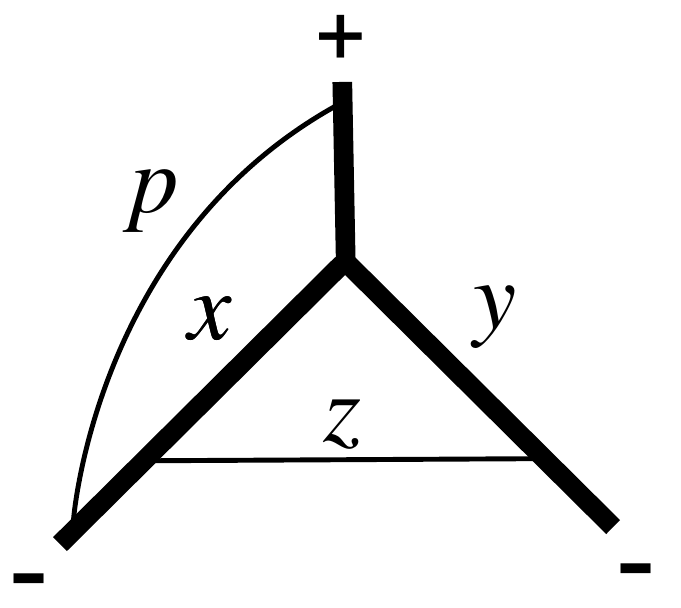}.$$ The output $z$ represents {\sc xor}. The tube $p$ represents {\sc nor} because a cytoplasmic flow is directed via $p$ if both tubes $x$ and $y$ are blocked. More complicated gates and memory devices can be found in~\cite{adamatzky2014slimefluidic}. 

\subsection{Intra-cellular collision based computing\index{collision-based computing}}

The paradigm of a collision-based computing originates from the computational universality of the Game of Life, Fredkin-Toffoli conservative logic and the billiard-ball model with its cellular-automaton implementation~\cite{adamatzky2002CBC}. A collision-based computer employs mobile localisations, e.g. gliders in Conway's Game of Life cellular automata, to represent quanta of information in active non-linear media. Information values, e.g. truth values of logical variables, are given by either the absence or presence of the localizations or by other parameters such as direction or velocity. The localizations travel in space and collide with each other. The results of the collisions are interpreted as computation. Physarum ballistic gates (Sect.~\ref{ballistic}) are also based on collisions, or interactions, between active growing zones of Physarum. However, signals {\emph per se}, represented by growing body of Physarum, are not localised. Intra-cellular vesicles --- up 100 nm droplets of liquid encapsulated by a lipid bilayer membrane --- could be convenient representations of localised signals.  

In \cite{mayne2015computing} we outlined pathways towards collision-based computing with vesicles inside the Physarum cell. The vesicles travel along actin and tubulin network and collide with each other. The colliding vesicles may reflect, fuse or annihilate. The vesicles reflect in over half of the collisions observed. The vesicles fuse in one of seven collisions. The vesicles annihilate and unload their cargo in one of ten collisions. The vesicles becomes paired and travel as a single object in one of ten collisions.  Based on the experimental observations, we derive  soft spheres collision~\cite{margolus2002universal} gates and  also gates {\sc not} and {\sc fan-out}.

\subsection{Kolmogorov-Uspensky machine\index{Kolmogorov-Uspensky machine}}

In 1950s Kolmogorov outlined a concept of an algorithmic process, an abstract machine, defined on a dynamically changing graph~\cite{kolmogorov1953concept}. The structure later became known as  Kolmogorov-Uspensky machine~\cite{kolmogorov1958definition}. The machine is a computational process on a finite undirected connected graph with distinctly labelled nodes~\cite{gurevich1988kolmogorov}. A computational process travels on the graph, activates nodes and removes or adds edges. A program  for the machine specifies how to replace the neighbourhood of an active node with a new neighbourhood, depending on the labels of edges connected to the active node and the labels of the nodes in proximity to the active node~\cite{blass2003algorithms}. The Kolmogorov-Uspensky machine is more flexible than a Turning machine because it recognises in real time some predicates not recognisable in real time by the Turing machine~\cite{grigor1980kolmogoroff}, it is stronger than any model of computation that requires $\Omega(n)$ time to access its memory~\cite{cloteaux2006some,shvachko1991different}. ``Turing machines formalize computation as it is performed by a human. Kolmogorov-Uspensky machines formalize computation as it performed by a physical process.''~\cite{blass2003algorithms}. 

We implement the Kolmogorov-Uspensky machine in the Physarum as follows~\cite{adamatzky2007physarum}. Stationary nodes are represented by sources of nutrients. Dynamic nodes are assigned to branching site of the protoplasmic tubes. The stationary nodes are labelled by food colourings because Physarum exhibits an hierarchy of preferences to different colourings. An active zone in the storage graph is selected by inoculating the Physarum on one of the stationary nodes. An edge of the Kolmogorov-Uspensky machine is a protoplasmic tube connecting the nodes.

Program and data are represented by the spatial configuration of stationary nodes. Results of the computation over a stationary data node are represented by the configuration of dynamic nodes and edges. The initial state of a Physarum machine (Physarum implementation of Kolmogorov-Uspensky machine) includes part of an input string (the part which represents the position of Physarum relative to stationary nodes), an empty output string, the current instruction in the program and the storage structure consisting of one isolated node. The physical graph structure developed by Physarum is the result of its computation. The Physarum machine halts when all data nodes are utilised. At every step of the computation there is an active node and an active zone (nodes neighbouring the active node). The active zone has a limited complexity: all elements of the zone are connected by some chain of edges to the initial node. The size of the active zone may vary depending on the computational task.  In the  Physarum machine, an active node is a trigger of contraction/excitation waves, which spread all over the plasmodium tree and cause pseudopodia to propagate, the shape to change and protoplasmic veins to annihilate. The active zone comprises stationary and/or dynamic nodes connected to an active node with tubes of protoplasm. Instructions of the Physarum machine are  {\sc input}, {\sc output}, {\sc go}, {\sc halt},  {\sc add node}, {\sc remove node}, {\sc add edge}, {\sc remove edge}, {\sc if}.  The  {\sc input} is done via distribution of sources of nutrients. The {\sc output} is recorded optically.  The {\sc set} instruction causes pointers to redirect. It is realised by placing a fresh nutrient source in the experimental container. When a new node is created, all pointers from the old node point to the new node~\cite{adamatzky2007physarum}.

\section{Electronics}

\subsection{Wires\index{wire}}

Protoplasmic tubes of the Physarum are conductive and therefore can be used as wires. A resistivity of Physarum protoplasmic tubes is of the same rank as resistivity of a cardiac and skeletal muscles of dogs and humans~\cite{geddes1967specific}. By using 1-5~cm protoplasmic tube as a wire we can light up a LED, and keep it illuminated  for days. We can power up a piezo audio transducer~\cite{adamatzky2013physarumwire} using the Physarum wire. Due to high, comparing to conventional conductors, resistivity of the Physarum we must apply a high voltage to power loads. For example, to operate the LED array we should apply 10 V and 3.9~$\mu$A direct current. To produce a 30dB sound with buzzer we need to apply 8 V.   

The Physarum wires are self-growing. To connect two pins of a circuit with a wire we inoculate Physarum at one pin and place a source of attractants near another pin. The Physarum grows a protoplasmic tube connecting two pins. The Physarum propagates well, and relatively fast, 1-5~mm/h, on a bare surface of electronic boards. Growing Physarum circuits can be controlled by  white and blue  light, chemical and thermal gradients, and electrical fields. Physarum wires can be robustly routed with a wide range of organic volatiles~\cite{de2014routing}.

Physarum wires can self-repair  after a substantial damage.  After part of a protoplasmic tube is removed (1-2~mm segment) the tube restores its integrity in six to nine hours.  Typically, a cytoplasm from cut-open ends spills out on a substrate. Each spilling of cytoplasm becomes covered by a cell wall and starts growing.  In few hours growing parts of the tube meet with each other and merge. Restoration of tubes conductivity was confirmed by electrical measurements~\cite{adamatzky2013physarumwire}.

Physarum wire can transfer analogue signals below 19~KHz without distortion~\cite{whiting2015transfer}. Physarum wires perform well in digital communication systems. For example, the protoplasmic tubes are used to establish communication between Arduino Mega and Digital 3-axis compass using I2C protocol. Valid magnometric data was confirmed by movement/rotation of the manometer and subsequent change in data receive alone the Physarum wire~\cite{whiting2015transfer}.   

Physarum wires are not immortal. To make them last longer we cover them with conducting organic polymer polypyrrole~\cite{de2015conducting}. A localised section of protoplasmic tube is treated with ferric chloride and exposed to vapour of pyrrole monomer. A 1~cm section of the treated tube has resistance 100 kOhm. The treatment is selective therefore we can produce live and functionalised Physarum wires in the same computing circuit.

\subsection{Low pass filter\index{low-pass filter}}

Physarum protoplasmic tubes are conductors. Do they modify analog or digital signals passing through them? 
Signal propagation in Physarum's protoplasmic tubes was tested using a  frequency response network 
analyser~\cite{whiting2015transfer}. Sinusoidal voltage waveforms are sent via the protoplasmic tubes at frequencies from 10Hz to 4MHz. The transfer function of voltage waveform passing through the tube is frequency dependent. 
Signals of higher frequencies are dramatically reduced in strength while low frequency waveforms remain largely unaffected. The magnitude-frequency profile matches a low pass filter. In most cases there was some attenuation of the voltage through the wire at the pass-band frequency range with a mean attenuation of -6dB. The cut off frequency was defined as -3dB from the pass-band magnitude; the mean cut off frequency was 19KHz. The phase-frequency response showed the 45 degree phase shift to be aligned very closely to the cut off frequency in all protoplasmic tubes~\cite{whiting2015transfer}.

\subsection{Oscillators\index{oscillator}}

An electronic oscillator is a device that produces periodic electronic signal. Electronic Physarum chips need oscillators to have a source of regularly spaced pulses. The experimental electronic oscillator, which converts direct current to alternating current signal, is made with Physarum as follows~\cite{adamatzky2014slimeoscillator}.  We made two electrodes setup --- Physarum spans two electrodes with a single protoplasmic tube, apply direct current potential in a range 2~V to 15~V and measure electrical potential difference between the electrodes.  When we apply an electrical potential to a protoplasmic tube we observe oscillations of the output electrical potential. The oscillations of the output potential are caused by periodic changes in resistance of the protoplasmic tube connecting the electrodes. Average resistance of a 10~mm protoplasmic tube is 3~MOhm.  Resistance of the protoplasmic tube exhibits oscillatory behaviour  with highly pronounced dominating frequency 0.014~Hz.  The resistance oscillations have average amplitude 0.59~MOhm, minimum amplitude of  resistance oscillations observed was 0.11~MOhm and maximum amplitude 1~MOhm. Oscillations in resistance observed are due to peristaltic contractions of the protoplasmic tube~\cite{sun2009single}. Average output potential and average amplitude of output potential oscillations grow linearly with the increase of the input potential. Frequency of oscillations remains almost constant. Physarum oscillator produces the same frequency oscillations at 2~V and 15~V applied potential.   A ratio of average amplitude of output potential oscillations to average output potential decreases by a power low with increase of input potential.

\subsection{Tactile sensor\index{tactile sensor}\index{sensor!tactile}}
\label{tactile}

A tactile sensor is a device that responds to a physical contact between the device and an object. 
When a segment of a glass capillary is placed across protoplasmic tube, which spans reference and recording electrodes,  Physarum demonstrates two types of responses to application of this load:   an immediate response with a high-amplitude impulse  and a prolonged response with changes in its oscillation pattern~\cite{adamatzky2013slime}.
The immediate response is a high-amplitude spike: its amplitude is 12.33~mV and its duration is 150~sec. The prolonged response is an envelop of increased amplitude oscillations.  For example, an average amplitude of oscillations before stimulation is 2.3~mV and duration  of each wave was 120~sec. The amplitude of waves in the prolonged response to stimulation is 5.29~mV  with a period of a wave increased to 124~sec. Tactile sensor developed in~ \cite{adamatzky2013slime} is non-reusable. While the load rests on the protoplasmic tube Physarum starts colonising the load. Removal of the load damages the protoplasmic tube. To rectify this deficiency we designed Physarum tactile bristle~\cite{adamatzky2014tactile}.

 To make a tactile bristle with Physarum we stuck a bristle in the agar blob on the recording electrode~\cite{adamatzky2014tactile}. In a couple of days after inoculation of Physarum to an agar blob on a references electrode the Physarum propagates to and colonises agar blob on a recording electrode.   Physarum climbs up the bristle and occupies one third to a half of the bristle's length.  The sensor works by deflecting the bristle. A sensed object does not come into direct contact with Physarum but only with a tip of the bristle not-colonised by Physarum. A typical response of Physarum to deflection of the bristle  is comprised of an immediate response ---- a high-amplitude impulse, and a prolonged response. High-amplitude impulse is always well pronounced, prolonged response oscillations can sometimes be distorted by other factors, e.g. growing branches of a protoplasmic tube or additional strands of plasmodium propagating between the agar blobs. Responses are repeatable not only in different experiments but also during several rounds of stimulation in the same experiment~\cite{adamatzky2014tactile}.

\subsection{Colour sensor\index{colour sensor}\index{sensor!colour}}
\label{colour}

A colour sensor is a device that gives wavelength-dependent response when illuminated. 
Physarum is photo-sensitive. It changes pattern of electrical potential oscillatory activity when illuminated~ \cite{block1981blue, wohlfarth1981pathway}. Moreover the Physarum distinguishes the colour of illumination~\cite{adamatzky2013towards}. We placed Physarum between two electrodes and illuminated it with red, green, blue or white light.  We also illuminated Physarum with white light via transparent lens.  Amount of light on the blob was 80-120 LUX for each colour.  We say the Physarum recognises a colour of the light  if it reacts to illumination with the colour by a unique changes in amplitude and periods of oscillatory activity.  We found that Physarum recognises when red and blue light are switched on and when red light is switched off.  Red and blue illuminations decrease frequency of oscillations. Red light increases amplitude of oscillations but blue light decreases the amplitude. Physarum does not differentiate between green and white lights.  Switching off red light leads to increase of period and decrease of amplitude of oscillations~\cite{adamatzky2013towards}.

\subsection{Chemical sensor\index{chemical sensor}\index{sensor!chemical}}
\label{chemical}

A chemical sensor is a device that gives a selective response when exposed to a target chemical substance. 
Physarum senses and responds to volatile aromatic substances~\cite{adamatzky2011attraction, adamatzky2012simulating, adamatzky2012physarum}.  We studied Physarum's binary 
preferences to various volatile chemicals~\cite{delacycostello2013assessing} and derived an experimental mapping between subset of chemoattractants and chemorepellents:  farnesene, tridecane, s(-)limonene, cis-3-hexenylacetate, geraniol, benzyl alcohol, linalool, nonanal and amplitude and frequency of electrical potential oscillation of Physarum~\cite{whiting2014towards}.  Physarum increases frequency of electrical potential oscillations when exposed to strongest attractants ---  farnesene, tridecane, s(-)limonene, cis-3-hexenylacetate. Exposure to repellents --- linlool, benzyl alcohol, nonanal --- leads to a decrease of oscillation frequency, and, for linlool and  benzyl alcohol,  increase of the oscillation amplitude. Physarum chemical sensor discriminates individual chemicals by changing amplitude and frequency of its electrical potential oscillations; it can detect chemical from a distance of several centimetres~\cite{whiting2014towards}.

\subsection{Memristors\index{memristor}}
\label{memristors}

A memristor is resistor with memory, which resistance depends on how much current had flown through the 
device~\cite{chua1971memristor, strukov2008missing}. Memristor is a material implication $\rightarrow$, a universal Boolean logical gate.   In laboratory experiments~\cite{gale2013slime}  we demonstrated that protoplasmic tubes of Physarum show current versus voltage profiles consistent with memristive system.  Experimental laboratory studies shown pronounced hysteresis and memristive effects exhibited by the slime mould~\cite{gale2013slime}.  Being a memristive element the slime mould's protoplasmic tube can also act a a low level sequential logic element~\cite{gale2013slime} operated with current spikes, or current transients. In such a device logical input bits are temporarily separated.   Memristive properties of the slime mould's protoplasmic tubes gives us a hope that a range of `classical' memristor-based neuromorphic architectures can be implemented with Physarum. Memristor is an analog of a synaptic connection~\cite{pershin2010experimental}. Being the living memristor each protoplasmic tube of Physarum may be seen as a synaptic element with memory, which state is modified depending on its pre-synaptic and post-synaptic activities. Therefore a network of Physarum's protoplasmic tubes is an associate memory network.
A memristor can be also made from Physarum bio-organic electrochemical transistor (Sect.~\ref{transistor}) by removing a drain electrode~\cite{cifarelli2014non}.

\subsection{Schottky diodes\index{Schottky diode}}

A diode is a two-terminal passive non-linear device which conducts mainly in one direction. The diodes are used as current rectifiers to change alternating current to direct current.  A forward voltage drop is a difference between electrical potentials of anode and cathode. A Schottky diode is a diode which has low forward voltage drop,  comparing to other families of diodes. A device showing some resemblance to Schottky diode  can be made of Physarum~\cite{cifarelli2014non}. In this device, the Physarum spans two asymmetrical junction electrodes --- gold and indium. Cyclic voltage-current characteristics are measured. The measurements reveal  suppression of conductivity for low voltage values in the direction of the positive bias and rectification features, pronounced more for the low values of the bias voltage~\cite{cifarelli2014non}. Physarum {\emph per se} is not a diode:  the rectifying properties emerge due to combination of features of the asymmetric electrode junction with electrochemical activity of the Physarum~\cite{cifarelli2014non}.

\subsection{Voltage divider\index{voltage divider}}

Voltage divider is a circuit that produces a given fraction of an input voltage as an output voltage~\cite{horowitz1989art}.  A 10~mm protoplasmic tube acts as a simple divider: the output voltage is 
c. 0.9 of the input voltages, voltages tested up to 20 V~\cite{adamatzky2013physarumwire}.  
A typical voltage divider has two resistances, represented by two protoplasmic tubes. The Physarum 
divider produces output with 12\% accuracy. This error is linear and might be due to differences in the protoplasmic tubes' resistances. Resistors in Physarum voltage divider can be made adjustable by applying illumination, heat, or loading the protoplasmic tubes with functional nano particles~\cite{mayne2015toward, whiting2015transfer}.  For example, loading Physarum with magnetite lower resistance of protoplasmic tube to 10-20 KOhm, making it compatible by value with common resistors~\cite{mayne2015toward}. One of the tube constituting the divider can be transformed to a potentiometer by making an output electrode a conductive micro needle~\cite{whiting2015transfer}.

\subsection{Transistors\index{transistor}}
\label{transistor}

A transistor is a three-terminal active device that power amplifies input signal. The additional power comes from an external source of power.  An organic  electrochemical transistor is a semiconducting polymer channel in contact with an electrolyte. Its functioning is based on the reversible doping of the polymer channel.  A hybrid Physarum bio-organic electrochemical transistor is made by interfacing an organic semiconductor, poly-3,4-ethylenedioxythiophene doped with poly-styrene sulfonate, with the Physarum~\cite{tarabella2015hybrid}. Physarum is used instead of electrolyte. Electrical measurements in three-terminal mode uncover characteristics similar to transistor operations. The device operates in a depletion mode similarly to standard  electrolyte-gated transistors. The Physarum transistor works well with platinum, golden and silver electrodes. If the drain electrode is removed and the device becomes two-terminal, it  exhibits cyclic voltage-current characteristics similar to memristors~\cite{tarabella2015hybrid}.

\subsection{Thermic switch}
\label{thermicswitch}

A thermistor is resistor which resistance changes depending on its temperature. When we heat a Physarum protoplasmic tube up to 40\textsuperscript{o}C it is resistance increases thousand times~\cite{walter2015hybrid}.  The temperature-induced increase of Physarum resistance is a threshold-wise. This is why the Physarum is not a thermistor but a thermic switch.  The Physarum thermic switches are reusable.  The full duty cycle, from the heat response to reforming, is  tens of minutes. The Physarum thermic switches are successfully tested in hybrid circuits implementing analog summators, 
{\sc and} and {\sc nand} gates, and cascades of the gates into a Flip-Flop latch.  The circuits performed well in multiple duty cycles with the same setups,  producing reproducible results from one duty cycle to another~\cite{walter2015hybrid}.

\section{Robotcs}

\subsection{Robot controllers\index{robot controller}}

Physarum responds to stimuli by changing pattern of its electrical potential oscillations and cytoplasm shuttling. 
By interfacing the Physarum with actuators we can make the slime mould controller for robots. Two prototypes of 
such robotic controllers are made: controller for a hexapod robot~\cite{tsuda2006robot} and controller for a robotic android head~\cite{galeAndroid}. 

A Physarum controller for hexapod robot is made of a star shaped template~\cite{tsuda2006robot}. It has six circular wells connected by the channels that meet at a single point. The Physarum grows inside the template. Physarum in each well acts as cytoplasm shuttle streaming oscillators. The Physarum oscillators in the wells coupled via Physarum body colonising the channels between the wells. Blue light used as a stimulus. The shuttle streaming of cytoplasm is measured via light absorbance. Oscillations of shuttle streaming in the wells, as a response to a stimulation with light, modulate phase and frequency of the robot legs's movement and cause the robot to change its direction of movement.  

An electrical activity of Physarum in response to stimulations is converted to affective state in the design of Physarum emotional controller~\cite{galeAndroid}. A Physarum is inoculated on a multi-electrode array and stimulated with nutrients (attractant) and light (repellent). Extracellular electrical potential is recorded. The recorded data is split into chunks.  We employed a circumplex model of affect, where emotions are plotted in two-dimensions determined by the  polarity and arousal level. The chunks are assigned polarity. Potential recorded during stimulation with attractant is given positive polarity. Data obtained during illumination of Physarum is assigned  negative polarity. A level of arousal is proportional to amplitude of the electrical potential. Emotions are assigned to the data chunks, based on the polarity and  the arousal of chunks and fed into an android robot. The data activate the motors placed in the positions matching sites of real muscles in a human face. Actuation of the motors causes movements of an artificial skin. 
The movements are expressed as affective facial expression of  the android~\cite{galeAndroid}. 

\subsection{Actuators\index{actuator}}

Physarum contracts its body in a phase with oscillation of calcium waves. By placing a column of water on one side of the Physarum dumbbell shape we calculated that the Physarum weighting 5~mg, can lift up a load 36 times heavier than its 
own weight~\cite{tsuda2012towards}.   In~\cite{adamatzky2010physarumboats} we studied how the Physarum can manipulate on a water surface. To make the Physarum propelling a `boat' we take  a small piece of a plastic foam, inoculate the Physarum on the foam and place this floater on a water surface. When Physarum is illuminated, it increases its peristaltic which transferred into a movement of the boat. If we allow Physarum to develop its protoplasmic tubes outside the float, on the water surface, then periodic contractions of tubes, stimulated by light, will make the Physarum boat propel away from the source of light. Another way to move the floater, is to place a stationary floater (anchor) with an attractants nearby the Physarum boat. Then Physarum develops protoplasmic tree towards the attractants and colonises the anchorr. Then Physarum straightens its tubes thus pulling the boat towards the anchor~\cite{adamatzky2010physarumboats}.

\subsection{Nervous system\index{nervous system}}

The Physarum senses tactile, chemical  and optical stimuli and converts the stimuli into characteristic patterns of its electrical  potential oscillations. The electrical responses to stimuli may propagate along protoplasmic tubes 
for distances exceeding tens of centimetres, like impulses in neural pathways do. The Physarum makes
decision about its propagation direction based on  information fusion from thousands of spatially extended 
protoplasmic loci, similarly to a  neurone collecting information from its dendritic tree.  When growing on a non-nutrient substrate Physarum develops shapes resembling body of a single neuron.  It looks like neuron --- can it be behave as  one? In~\cite{adamatzky2015alife} we  speculate on whether an  alternative ---- would-be --- nervous systems can be developed and practically implemented from the slime mould.   We uncover analogies between the slime mould and neurons, and demonstrate that the slime mould can play a role of primitive mechanoreceptors, photoreceptors, chemoreceptors; we also show how the Physarum neural pathways develop~\cite{adamatzky2015alife}. 

 Physarum neural networks do not have synapses represented as discrete structural elements. Synapse-like morphological 
contacts could not be formed. When two pieces of Physarum are inoculated at a distance form each other, they start exploring space around them and form branching networks of protoplasmic tubes. When two networks, grown from different sites of inoculation come into contact they usually fuse forming a single united network. However, there is
a functional analog of synapses and an intrinsic feature of Physarum protoplasmic tubes which makes literally any loci of Physarum network a synapse. This is a memristive property (Sect.~\ref{memristors}).

We explored the analogy between behaviour of neuron growth cones and Physarum active growing zones~\cite{adamatzky2015alife}. To test if Physarum can develop information pathways we conducted several experiments on one to one scale models of human skull and brain.  We used real scale models for the following reasons. First, to show that information pathways made of protoplasmic tubes can be tens of centimetres length and thus match lengths of neural pathways. Second, to demonstrate that --- when propagating inside human skull --- the plasmodium follows general anatomical  trajectories of ocular and olfactory nerves.  We found that morphology of information pathways developed by Physarum on a human skull matches well anatomy of the real nervous pathways. Impressive results we also obtained in imitation of sensorial innervation of the front scalp: we inoculated Physarum on the frontal bone above 
glabella and placed few oat flakes on the pariental bone. In two days Physarum  developed an extensively branching tree of protoplasmic tubes. The tree spanned substantial part of the frontal lobe, even covering its lateral parts, crossed coronal sutura and developed actively branching growing zones moving towards the target site on the parietal bone~\cite{adamatzky2015alife}.

\subsection{Illusions\index{illusion}}

A configuration of three Pac-Man shapes positioned at a vertices of an imaginary triangle and looking towards the centre of the triangle is a  famous illusory contour~\cite{kanizsa1976subjective}.  When we look at this configuration of the shapes we are getting an impression of a white triangle defined by the Pac-Man shapes.  The illusory contour disappears when the Pac-Man shapes are facing away from each other. Physarum shows tendency to `experience' the same illusion as humans do~\cite{tani2014kanizsa}.  The Pac-Man shapes are made of a nutrient rich agar and placed on a non-nutrient agar. The Physarum is inoculated in the centre of each Pac-Man. The Physarum develops protoplasmic networks spanning the configuration of Pac-Man shapes.  When spanning the  configuration, when Pac-Man shapes are facing each other, the Physarum  formed a network matching a contour of the illusory triangle in 4/5 of experiments~\cite{tani2014kanizsa}. In the scenarios of away facing Pan-Man shapes the Physarum matched the illusory triangle only in the half of experiments and constructed spanning tree and other graphs in another half of experiments. Conclusion was that Physarum shows behaviour which mimic illusory impressions of humans~\cite{tani2014kanizsa}. Exact mechanisms of such behaviour of Physarum are different from humans, however there may be some subtle analogies between how we visually scan pictures~\cite{yarbus1967eye} and how the Physarum perceives  its environment.  


\section{Energy production}

\subsection{Modulation of energy generation\index{energy generation}}

A microbial fuel cell is a biological electrochemical  device which uses micro-organisms to convert energy of organic substates into an electrical energy. The micro-organisms colonise electrodes, metabolise organic material and donate electrons to the electrodes~\cite{ieropoulos2005comparative}. A basic fuel cell is made of two electrodes and an ion selective membrane. Electrodes are made of folded carbon veil. Two scenarios are explored in~\cite{taylor2015physarum} for anode and cathode sites of Physarum inoculation.  When Physarum is housed on anode no significant difference between the test and the control fuel cells is observed. In experiments with Physarum inoculated on cathode a statistically significant power increase is observed. A peak open circuit voltage of Physarum fuel cell and control fuel cell was around 0.6V~\cite{taylor2015physarum}.  When 9.4kOhm external load is connected,  the voltage drops to 0.2V in the control fuel cell and 0.25V in the Physarum fuel cell. During 7 days of experiments power generated by the Physarum fuel cell was 12.5--15~$\mu$W and power generated by the control fuel cell 10--11V~\cite{taylor2015physarum}. 

\subsection{Biodiesel production\index{biodiesel}}

This is not about computation, sensing, actuation or analog physical modelling. But we included this section because future unconventional devices will need energy. Biodiesel is a  liquid fuel made of vegetable oil or animal fat and used in compression-ignition engines. It is produced by a conversion of a carboxylic acid ester into a different carboxylic acid ester by reacting lipids with an alcohol producing fatty acid esters~\cite{ma1999biodiesel}. Biodiesel is advantageous to fossil fuels because it is biodegradable and has low toxicity. However, biodiesel is expensive to produce: cost of raw material is 2/3 of production costs.  Slime moulds have relatively high concentration of lipids in their bodies.  Thus they could be a good alternative to existing technologies of bio-mass production. \emph {P. polycephalum} is found to be a champion, amongst \emph{Myxomecetes}, in lipid production~\cite{tran2012evaluating}. Physarum produces equivalent of 65~g of dry biomass and over 7~g of lipids per litre of culture medium in four days~\cite{tran2012evaluating}. The biomass production rate  exceeds that of algae~\cite{chisti2007biodiesel}. Tran et al \cite{tran2015evaluation}  show that cultivation of Physarum on 37.5 g per litre rice brans yields 7.5 g dry biomass and 0.9 g lipid in five days.

\section{Arts}
 
\subsection{Music generation\index{music generation}}

Physarum expresses its physiological states in patterns of its electrical potential oscillators~\cite{adamatzky2011electrical}. In \cite{adamatzky2009music}  5-10 days recordings were processed and converted to sound track by mapping parameters of electrical potential oscillation to pitch, attack and duration of tones. First ever sound track was produced from Physarum's electrical activity in ~\cite{adamatzky2009music}. The music reflected a physiological transition of Physarum from a comfortable foraging state to a state of active search for disappearing nutrients to decision making state to transformation to sclerotium. When the track played to auditorium at various presentations, majority of people were getting a feeling of a dramatic development in the  `life of Physarum'.  Later the transformation of Physarum activity recording into sounds has been taken at a more professional, from music point of view, level in~\cite{miranda2011sounds}. There electrical activity of Physarum was converted to parameters of sinusoidal oscillators;  the rhythmic behaviour of Physarum was shown to produce different timbres.

 Memristive properties of Physarum (Sec.~\ref{memristors}) are used to generate  musical responses in~\cite{mirandamemristors}.  A vocabulary notes are assigned voltage values.  The Physarum current-voltage response to electrical stimulation is recorded.  Discrete voltages are converted to notes. The notes are fed into  a MIDI keyboard. 
During interactive  music performance between human composer and Physarum, a feedback to the Physarum is implemented. Parts of well-know melody: Elgar's Nimrod and  Beethoven's F\"{u}r Elise were generated for live performances with Physarum in ~\cite{mirandamemristors}.

\subsection{Modelling creativity\index{creativity}}

Creativity is manifested by divergent thinking and lack of lateral inhibition, making remote associations between ideas and concepts, switching  between ideation, generating ideas of actualities, risk taking, nonconformity~\cite{kuszewski2009genetics}.  In~\cite{adamatzky2013creativity} we show how to use live Physarum to analyse many scenarios of an individual creativity genesis. The divergent thinking is expressed by Physarum in its simultaneous reaction to several sources of attractants and repellents, and parallel implementation of sensorial fusion. Functional non-conformity  of Physarum is manifested in its self-avoidance. Activity in both hemispheres and exchange of activities between hemispheres are considered to be attributes of human creativity~\cite{kuszewski2009genetics}.  In Physarum the hemispheres' activity  is represented by simultaneous oscillatory activity, with biochemical oscillators located to distant parts of Physarum body and oscillating with different frequencies and amplitudes. Interaction between `hemispheres' is instantiated by waves of calcium waves and waves of contractile activity propagating along protoplasmic tubes.

Cognitive control of divergent thinking is a requisite of creativity\cite{kuszewski2009genetics}. A person with extremely divergent thinking who is unable to control these associations would be potentially classified as mentally ill. However, those who can fit their high schizotypy (a range of personality characteristics ranging from normal to schizophrenia) traits into rigorous cognitive frameworks may be classified as gifted or even genius.  Thus creativity could be positioned together with autism and schizophrenia in the same phase space. Physarum imitates cognitive control by tuning regularity of its protoplasmic network.  A degree of branching  of a Physarum network may be considered as a representation of a degree of schizotypy. Then `mathematical savant' slime mould grows a low branching highly symmetrical protoplasmic networks  and severely `autistic' Physarum develops highly asymmetric low branching networks~\cite{adamatzky2013creativity}.

\section{Things inspired by Physarum but never done with a real one}

\begin{enumerate}
\item A non-quantum implementation of Shor's factorisation \index{Shor's factorisation} algorithm~\cite{blakey2014towards} is inspired by Physarum ability to retain the time-periods of stimulation, the anticipatory behaviour~\cite{saigusa2008amoebae} 
\item Single electron circuit \index{single electron circuit} solving maze problem~\cite{shinde2014design} is based on a cellular automaton imitation of Physarum.  
\item Particle based model of Physarum approximates moving average and low-pass filters \index{low-pass filter} in one-dimensional data sets and spatial computation of splines in two-dimensional data set~\cite{jones2014material}.
\item Physarum logic \index{Physarum logic} is developed by interpreting basic features of the Physarum foraging behaviour in terms of process calculi and spatial logic without modal operators~\cite{schumann2011physarum}.
\item Soft amoeboid \index{amoeboid robot} robots~\cite{piovanelli2012bio} ---- models of coupled-oscillator-based \index{coupled oscillators} robots~\cite{umedachi2013fluid} are based on general principles of Physarum behaviour, especially coordination of distant parts of its cell. 
\item Physarum concurrent games \index{concurrent game} are proposed in  \cite{schumann2014bio}.  In these games rules can change, players update their strategies and actions, resistance points are reduced to payoffs.  At the time these games were proposed, we were  unaware about the lethal reaction followed a  fusion between Physarum of two different strains~\cite{carlile1972lethal}: a degree of lethality depends on the position and size of invasion between strains, which supports the ideas developed in  \cite{schumann2014bio}.
\item  Physarum is interfaced with a \index{field-programmable} field-programmable array in \cite{mayne2015towardsFPGA}. The hybrid system performs predefined arithmetic operations derived by digital recognition of membrane potential oscillations. 
\item A Physarum-inspired algorithm for solution of the Steiner tree \index{Steiner tree} \index{tree!Steiner} problem is proposed in \cite{liu2015physarum} and applied to optimise the minimal exposure problem and worst-case coverage in wireless sensor networks~\cite{liu2015physarum, tsompanas2015cellular}. 
\item   An abstract implementation of reversible logical gates \index{reversible logical gate} \index{logical gate!reversible} with Physarum is proposed in \cite{schumann2015conventional}. 
\item Mechanisms of Physarum foraging behaviour are employed in robotics algorithm for simultaneous localisation and mapping~\cite{kalogeiton2014hey}.
\item A range of algorithms for network optimisation \index{network optimisation} is derived from a model of Physarum shortest path \index{shortest path} formation~\cite{tero2006physarum, bonifaci2012physarum}. Most of these algorithms are based on a feedback between traffic throughout a tube and the tube's capacity.  They include dynamical shortest path algorithms~\cite{zhang2014improved}, optimal communication paths in wireless sensor networks~\cite{zhang2015physarumPPL, dourvas2015hardware}, supply \index{supply chain design} chains design~\cite{zhang2015physarumPPL},  shortest path tree problem~\cite{zhang2015physarum},  design of fault \index{fault tolerant graph} tolerant graphs \cite{becker2015evaluating}, and multi-cast \index{multi-case routing} routing~\cite{liang2015new}. Behaviour of  Physarum solver \cite{tero2006physarum} was also applied to deriving a shortest path on \index{Riemann surface} Riemann surface~\cite{miyaji2008physarum}.
\item Physarum-inspired algorithm for learning Bayesian network \index{Bayesian network} structure from data is designed in \cite{schon2014physarum}. 
\item Attraction based two-input two-output gate realising {\sc and} and {\sc or} and three-input two-output gate realising conjunction of three inputs and negation of one input with disjunction of two other inputs are constructed in particle-based model of Physarum~\cite{jones2010towardsadder}; the gates are cascaded into a one-bit adder.
\item A nano-device aimed to solve Boolean satisfiability problem, inspired by optimisation of protoplasmic networks by Physarum, is designed in \cite{aono2012amoeba, aono2015amoeba}. The device works on fluctuations generated from thermal energy in nanowires, electrical Brownian ratchets~\index{Brownian ratchet}.
\item Solution of a the `exploration versus exploitation' dilemma by Physarum making a choice between colonising nutrients and escaping illuminated areas is used in a tug of war \index{tug of war} model~\cite{kim2010tug}: a parallel search of a space by collectives of  locally correlated agents and decision making in situations of uncertainty.  The ideas are developed further in a design of experimental device where a single photon solves the multi-armed bandit \index{multi-armed bandit} problem~\cite{naruse2015single}.
\end{enumerate}

 \section{Post coitum omne animal triste est}
  
Excitement of being able to make computing, sensing and actuating devices with live Physarum eventually fades down. 
Let us wake up now. Is there any real use of the slime mould?  Is it fast? No. Physarum is a very slow creature. It might take the Physarum several days to compute Voronoi diagram or a shortest path. Is it quick in responding to stimuli? Not really. Is it robust?  No two experiments are the same. Physarum always behaves differently. Results of a computation performed by Physarum are only valid when at least dozen of experiments are done cause a single trial might mean nothing. Are the Physarum computing devices reliable? Failure rate is near 30\%.  The slime mould is not a miraculous computing substrate. It is just a user-friendly non-demanding living creature which changes its form and shape to stay comfortable in the fields of attractants and repellents.  Why did we love Physarum then? Because by taking myriad of different shapes and exhibiting fantastically rich spectrum of electrical potential oscillations Physarum makes a unique fruitful material for interpretations of its behaviour in terms of purposeful transformation of data to results. The unconventional computing is an art of interpretation. Physarum feeds out phantasy and fuels our unconventional computing dreams. Will we ever see the Physarum in commercial computing or sensing devices? Not tomorrow. In no way Physarum can win over the silicon technology which has been optimised non-stop for decades and decades.  But a success depends on many factors, not just technological ones. Success is in finding a  vacant niche and flourishing there.  More likely applications of Physarum computers will be in disposable hybrid processing devices used for sensing and decision-making in environments and situations where speed does not matter but being energy efficient, adaptable and self-healing is important.

\printindex

\end{document}